\documentclass[fleqn,usenatbib]{mnras}

\usepackage{newtxtext,newtxmath}

\usepackage[T1]{fontenc}

\DeclareRobustCommand{\VAN}[3]{#2}
\let\VANthebibliography\thebibliography
\def\thebibliography{\DeclareRobustCommand{\VAN}[3]{##3}\VANthebibliography}

\usepackage{graphicx}	
\usepackage{amsmath}	

\title[Reionization in Large Hydrodynamic Simulations]{Computationally Efficient Reionization in a Large Hydrodynamic Galaxy Formation Simulation}

\author[James E. Davies et al.]{
James E. Davies$^{1}$\thanks{E-mail: james.davies@sns.it},
Simeon Bird$^{2}$,
Simon Mutch$^{3,4}$,
Yueying Ni$^{5,6}$,
Yu Feng$^{7}$,
Rupert Croft$^{6}$,
\newauthor
Tiziana Di Matteo$^{6}$,
and J. Stuart B. Wyithe$^{3,4}$,
\\
$^{1}$Scuola Normale Superiore, Piazza dei Cavalieri 7, I-56126 Pisa, Italy\\
$^{2}$Department of Physics \& Astronomy, University of California, Riverside, 900 University Ave., Riverside, CA 92521, USA\\
$^{3}$School of Physics, The University of Melbourne, Parkville, Victoria 3010, Australia\\
$^{4}$ARC Centre of Excellence for All Sky Astrophysics in 3 Dimensions (ASTRO 3D)\\
$^{5}$Harvard-Smithsonian Center for Astrophysics, 60 Garden Street, Cambridge, MA 02138, USA\\
$^{6}$McWilliams Center for Cosmology, Carnegie Mellon University, Pittsburgh PA, 15213\\
$^{7}$Berkeley Center for Cosmological Physics, University of California at Berkeley, Berkeley, CA, 94720, USA\\
}

\date{Accepted XXX. Received YYY; in original form ZZZ}

\pubyear{2023}

\begin{document}

\label{firstpage}
\pagerange{\pageref{firstpage}--\pageref{lastpage}}
\maketitle

\begin{abstract}
Accuracy in the topology and statistics of a simulated Epoch of Reionization (EoR) are vital to draw connections between observations and physical processes. While full radiative transfer models produce the most accurate reionization models, they are highly computationally expensive, and are infeasible for the largest cosmological simulations. Instead, large simulations often include EoR models that are pre-computed via the initial density field, or post-processed where feedback effects are ignored. We introduce \textsc{Astrid-ES}, a resimulation of the \textsc{Astrid} epoch of reionisation $20 > z > 5.5$ which includes an on-the-fly excursion-set reionization algorithm. \textsc{Astrid-ES} produces more accurate reionization histories without significantly impacting the computational time. This model directly utilises the star particles produced in the simulation to calculate the EoR history and includes a UV background which heats the gas particles after their reionization. We contrast the reionization topology and statistics in \textsc{Astrid-ES} with the previously employed parametric reionisation model, finding that in \textsc{Astrid-ES}, ionised regions are more correlated with galaxies, and the 21cm power-spectrum shows an increase in large scale power. We calculate the relation between the size of HII regions and the UV luminosity of the brightest galaxy within them. Prior to the overlap phase, we find a power-law fit of $\mathrm{log} (R) = -0.314 M_\mathrm{UV} - 2.550 \mathrm{log}(1+z) + 7.408$ with a standard deviation $\sigma_R < 0.15 \mathrm{dex}$ across all mass bins. We also examine the properties of halos throughout reionization, finding that while the properties of halos in the simulation are correlated with the redshift of reionisation, they are not greatly affected by reionisation itself.
\end{abstract}



\section{Introduction}
Constraining the epoch of reionization requires theoretical modelling which spans a vast range of physical scales, from the growth of large scale structure and formation of early galaxies (see e.g. \citet{Dayal18}), to the interactions between hydrogen atoms and the Lyman Continuum, Lyman-Werner, and X-ray photon backgrounds from these early sources, as well as the CMB (see e.g \citet{Furlanetto06}). As such there is always a trade-off between the level of detail in our physical models and their computational cost. More expensive models involve detailed prescriptions for the generation and propagation of ionising photons \citep{Kannan21,Ocvirk20,Lewis22}, allowing comparison with a wider range of observational data. On the other hand, more computationally cheap modelling frameworks use more approximate methods to simulate the connection between early ionising sources and the topology of reionisation \citep{2011MNRAS.411..955M,2016MNRAS.462..250M,Molaro18,Choudhury18,Hutter21}. As a result these models can be run on larger volumes, at higher resolutions, and over a wider range of model parameters with the same computational resources.

The most detailed models of reionization are radiative transfer models, which explicitly model photons within a simulation and follows their propogation through the IGM, accounting for absorption along the way \citep{Mihalas84}. They take into account the complex geometry and distributions of both ionising sources and absorbers. However the high dimensionality of the equations which discretise over position, time, angle and frequency, as well as the modelling of several chemical processes in the intergalactic medium make this approach highly computationally expensive.

\begin{figure*}
    \centering
    \includegraphics[width=\linewidth]{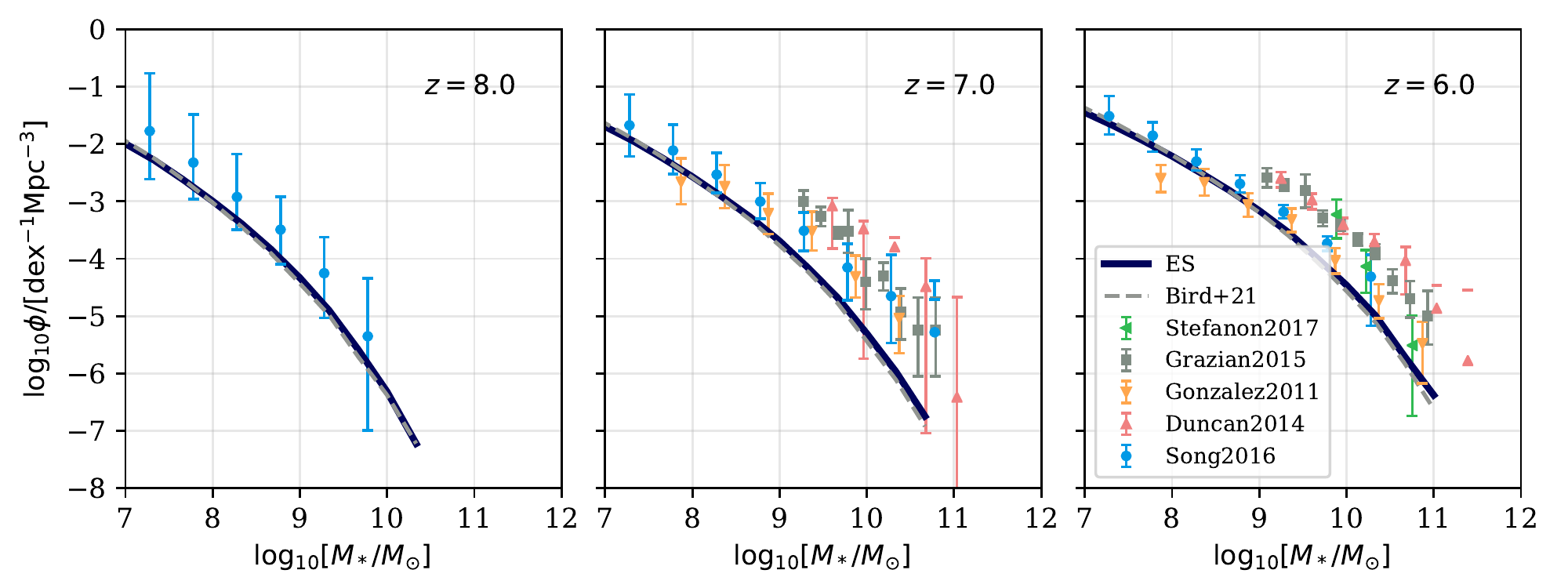}
    \caption{Galaxy stellar mass functions for \textsc{Astrid-ES} (solid lines), compared with observational data from \citet{Stefanon17,Grazian15,Gonzalez11,Duncan14}, and \citet{Song16}. We also show the stellar mass function from \textsc{Astrid} \citep{Bird21} as a grey dashed line, which is consistent with \textsc{Astrid-ES}, as the reionisation timing does not have a significant effect on the total stellar mass in most galaxies (see section \ref{sec:feedback}). As in \citet{Bird21}, we find that galaxy stellar masses are $\sim 1 \sigma$ below the bulk of observations at redshifts $7 \leq z \leq 8$.}
    \label{fig:gsmf}
\end{figure*}

The largest hydrodynamic simulations such as \textsc{Bluetides} \citep{YFeng16} and \textsc{Astrid} \citep{Bird21} are too large to utilise radiative transfer models whilst remaining computationally feasible, although recent works have coupled radiative transfer with hydrodynamic simulations up to a cube of side length $<70 h^{-1} \mathrm{cMpc}$ \citep{Kannan21,Ocvirk20}. Instead, these large simulations use parametric models which pre-determine reionization history according to the initial conditions of the simulation  \citep{Battaglia2013,Onorbe19}. While these methods were shown to reproduce certain large-scale characteristics of a radiative transfer simulation, the topology of reionization in these models is disconnected from the ionising sources in the simulation volume. There are also efficient post-processed radiative transfer models which calculate an inhomogeneous EoR history from the sources in a simulation, either ignoring inhomogeneous photoheating feedback onto the sources \citep{Keating18} or requiring a second hydrodynamic simulation to fully capture the effects of reionization on galaxies \citep{Puchwein23}.
 
The excursion-set reionization algorithm has been used extensively to produce realistic ionisation topologies in large volumes whilst being very computationally efficient, and occupies a middle ground between radiative transfer models and analytic models. The algorithm can be applied to an evolved density field in semi-numerical models \citep{2007ApJ...669..663M,2011MNRAS.411..955M,Choudhury18,Molaro18,Park19,Onorbe19} or discrete ionising sources in semi-analytic galaxy evolution models \citep{2016MNRAS.462..250M,Hutter18,Hutter21}. The radiation field is approximated by constructing grids of gas density and ionising emissivity, filtering them on a series of scales, and flagging cells as ionised in regions where the number of ionising photons generated by the stars within the filter scale is greater than the number of neutral hydrogen atoms. The large scale structure of reionization produced by similar models has been shown to compare favorably to radiative transfer models \citep{Hutter18,Hassan22}.

In this paper, we have adapted the \textsc{21cmFAST} \citep{2011MNRAS.411..955M} excursion-set algorithm to perform on-the-fly reionisation calculations within \textsc{MP-Gadget}. We take the distribution of gas and stellar particles as input, and direct its output of the UV background to the existing cooling model, which evolves the ionisation and thermal state of each particle. This algorithm allows us to study more realistic histories of reionization which are directly dependent on the sources of ionising photons within the simulation and feeds back onto those sources as the gas is heated. The excursion-set algorithm is computationally cheap compared to the hydrodynamics, such that the overall run-time of any simulation will be negligibly affected.

In section \ref{sec:code}, we briefly introduce the main features of the \textsc{MP-Gadget} codebase. In section \ref{sec:es}, we detail the excursion-set algorithm and ionising background model. In section \ref{sec:results}, we compare the reionization history of the excursion-set with the previous parametric model, and present new results regarding the correlation between galaxy properties and the EoR.

\section{MP-Gadget}\label{sec:code}
\textsc{MP-Gadget} uses a pressure-entropy formulation of Smoothed Particle Hydrodynamics \citep{Read10,Hopkins13}. This includes a multi-phase star formation model \citep{Springel03,Vogelsberger13}, cooling via radiative processes \citep{Katz96}, metal cooling \citep{Vogelsberger14}, the effects of molecular hydrogen on star formation \citep{Krumholz11}, a type II supernovae feedback model \citep{Okamoto10}, helium II reionization \citep{UptonSanderbeck20}, and a super-massive black hole treatment including dynamical friction \citep{DiMatteo05,Chen21}. The $250\mathrm{h^{-1}cMpc}$ \textsc{Astrid} simulation has been shown to produce UV luminosity functions, stellar mass functions and specific star formation rates consistent with observation in the redshift range $3<z<10$ \citep{Bird21}, as well as black hole mass functions in the redshift range $3<z<6$ \citep{Ni22}. Both \textsc{Astrid} and our resimulation of reionisation \textsc{Astrid-ES} introduced in this paper have a dark matter particle mass of $6.75 \times 10^6 h^{-1} M_\odot$ and a gas particle mass of $1.267 \times 10^6 h^{-1} M_\odot$. The Planck 2015 cosmology \citep{Planck2016} is used for each simulation $\lbrace \Omega_{\mathrm m}, \Omega_{\mathrm b}, \Omega_\Lambda, h, \sigma_8, n_{\mathrm s} \rbrace =\lbrace 0.3089, 0.0486, 0.6911, 0.6774, 0.816, 0.9667 \rbrace$.

\subsection{Parametric Reionization}
The reionization model in \textsc{MP-Gadget} used in the \textsc{Astrid} simulation is a parametric model outlined in \citet{Battaglia2013}. We expect the EoR to begin in the dense filaments of the cosmic web, as these areas are likely to contain both larger numbers of galaxies and brighter galaxies overall. \citet{Battaglia2013} drew spatial correlations between the large-scale density field and the expected reionization history by analysing the outputs of radiative transfer simulations. The result of this work is a Fourier-space filter that can be applied to the density fields of simulations in order to predict their reionization history without directly performing the expensive radiative transfer calculation.

\begin{figure}
    \centering
    \includegraphics[width=0.97\linewidth]{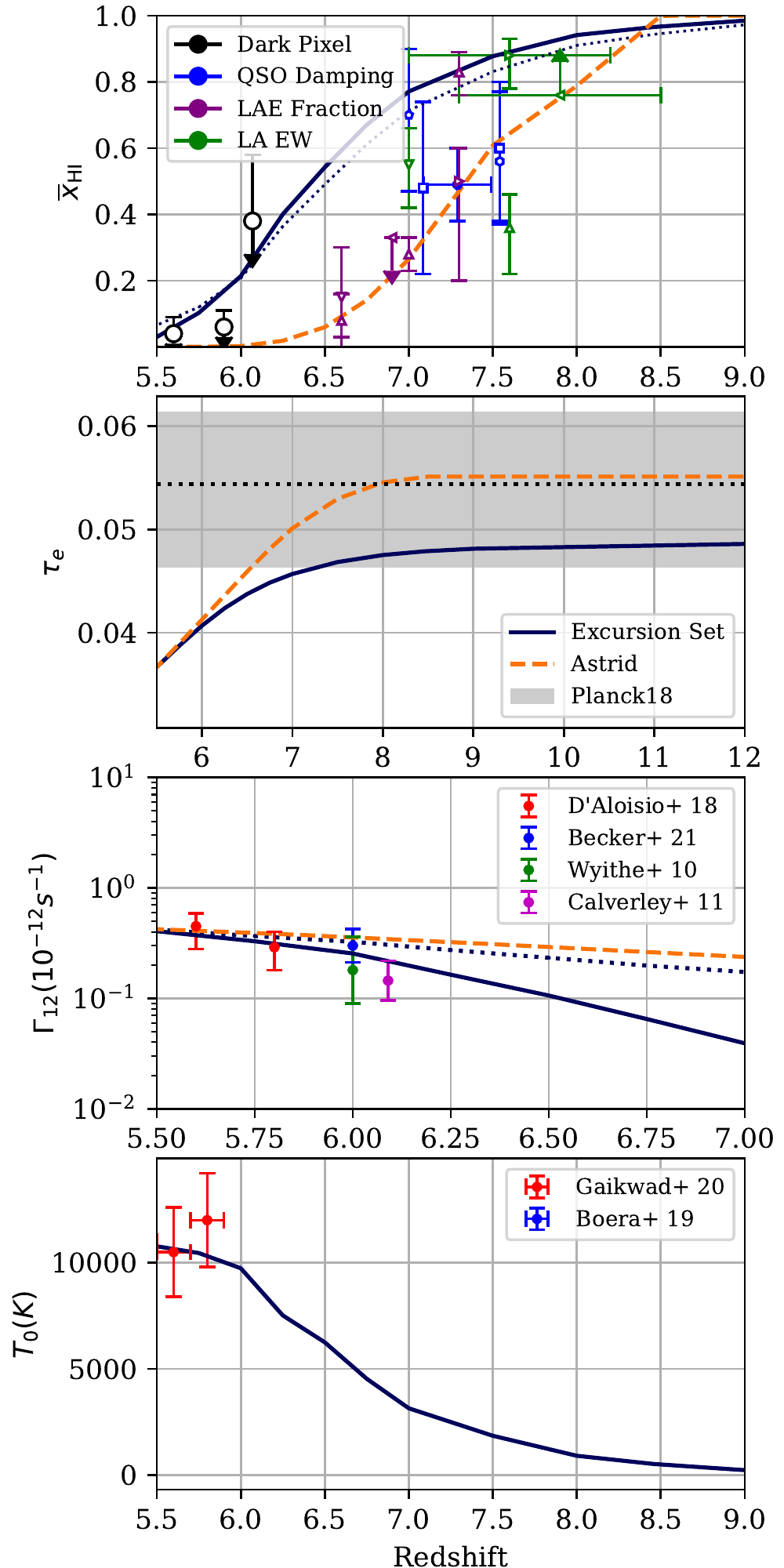}
    \caption{Top panel: Global reionization histories showing the volume (solid line) and mass (dotted line) weighted neutral fraction of \textsc{Astrid-ES}, as well as the volume-weighted fraction \textsc{Astrid} \citep{Bird21} (orange line). We compare with observations from Dark pixel QSO damping wing, lyman alpha emitter fraction, and equivalent width observations. \citep{McGreer15,Inoue18,Ouchi18,Morales21,Wold21,Hoag19,Mason19,Whitler20,Greig17,Davies18,Greig19,Wang20}. Second panel: Thomson optical depth of each simulation, compared with the Planck 2018 observation \citep{Planck2018}. Third panel: volume-weighted ionisation rate for all particles (solid line) and ionised particles (dotted line) in \textsc{Astrid-ES}, as well as the ionisation rate applied to ionised regions in \textsc{Astrid} \citet{FG19} (dashed orange line). Ionisation rates are compared with observations \citep{Wyithe11,Calverley11,DAloisio18,Becker21}. Bottom panel: Temperature at mean density in the excursion set model, compared with observations from \citet{Gaikwad20}. We do not show the temperature history of \textsc{Astrid}, since heating from the initial ionisation of a particle is not included.}.
    \label{fig:globaleor}
\end{figure}

In these models, the bias between overdensity and the redshift of reionization $b_{mz}$ is fit to a functional form:

\begin{equation}
    b_{mz}(k) = \frac{b_0}{(k + k/k_0)^\alpha}
\end{equation}
where $k_0$ and $\alpha$ characterises the scaling of the bias with Fourier mode $k$. These values were found to be $k_0 = 0.185 h^{-1} \mathrm{Mpc}$ and $\alpha = 0.564$. $b_0$ is the normalisation of this bias on large scales and is derived from the extended Press-Schecter formalism \citep{Bond91} $b_0 = \frac{1}{\delta_{\mathrm{crit}}} = 0.593$. The resulting bias is applied alongside a spherical tophat filter of radius 1 Mpc to the initial conditions of the simulation, which has been evolved to the desired median redshift of reionization via second order Lagrangian perturbation theory.
The resulting $z_\mathrm{reion}$ field is applied in the simulation by exposing particles within each cell with $z < z_\mathrm{reion}$ to a redshift-dependent UV background calculated by \citet{FG19}. Since the bias and reionization fields are pre-computed, this approach effectively adds no computational cost to the simulation. However, the model does not directly correlate the properties of the luminous sources in a simulation with its reionization history, instead relying on correlation with the underlying density field.

\begin{figure*}
    \centering
    \includegraphics[width=\linewidth]{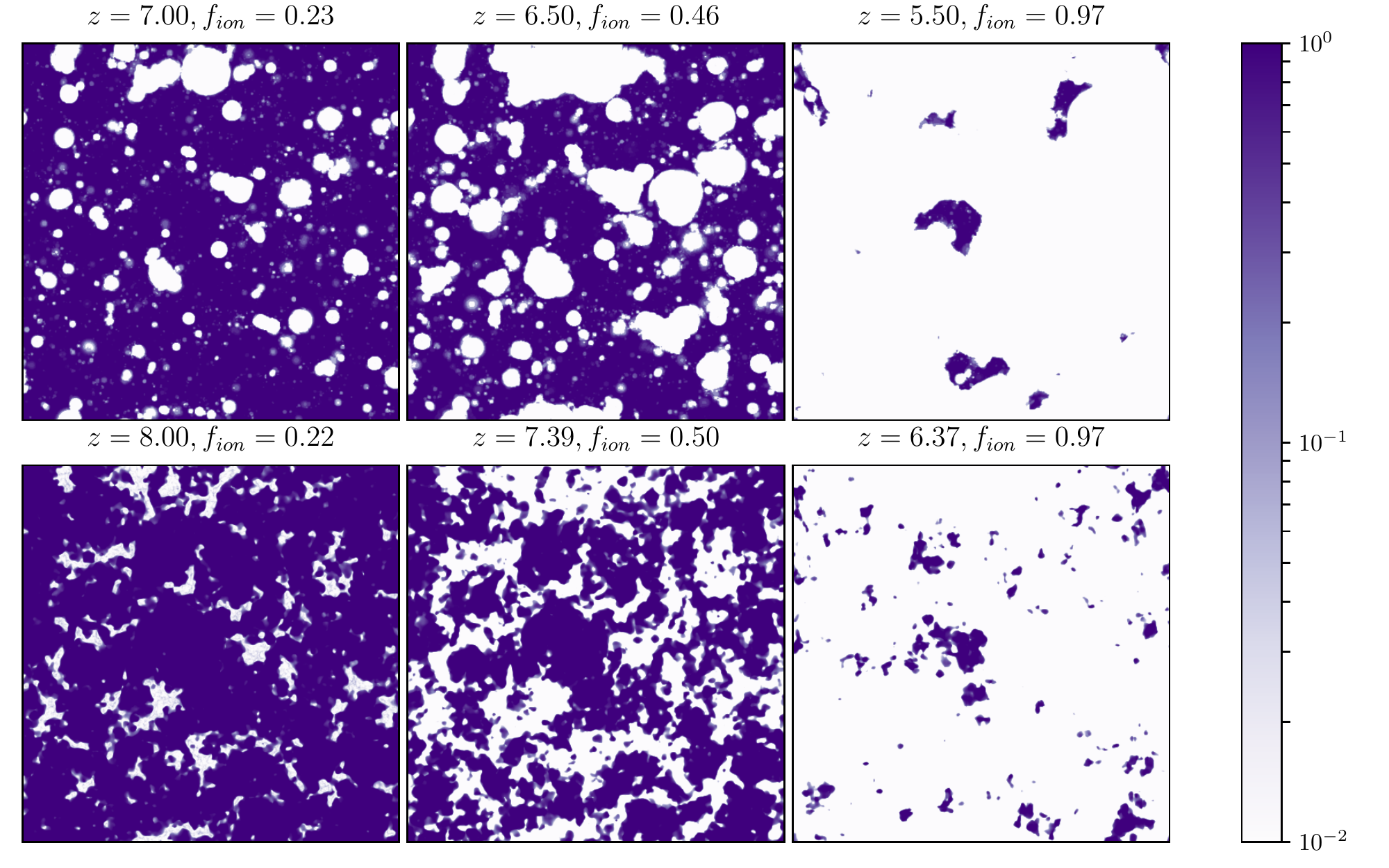}
    \caption{$250 \times 250 \times 5 h^{-1} \mathrm{Mpc}$ neutral hydrogen fraction slices in the \textsc{Astrid-ES} excursion set (top) and \textsc{Astrid} parametric (bottom) reionization models. To account for the differing reionization histories,  columns correspond to different stages of reionisation, with a fraction of ionised cells $Q_\mathrm{HII} \approx 0.22, 0.50, 0.97$ in each model. The ionisation structure in the \textsc{Astrid-ES} is composed of overlapping regions which are highly spherical, centred around bright galaxies. In \textsc{Astrid}, the ionisation structure closely follows the cosmic filaments of the underlying density structure.}
    \label{fig:reion50}
\end{figure*}

\section{Excursion-Set Algorithm}\label{sec:es}
The large volume ($250 h^{-1} \mathrm{cMpc}$ side-length) and high mass-resolution ($6.75 \times 10^6 h^{-1} M_\odot$ per dark matter particle) of \textsc{Astrid} means that a full radiative transfer model to estimate the UV background is far too computationally expensive. The excursion-set algorithm, however, is able to produce accurate reionization histories using the ionising sources within the simulation while requiring very little memory and computation time compared with the rest of the simulation. As a result it can be performed during the simulation run with very little additional cost.

Our algorithm for calculating the reionization history within the simulation is based on the version of \textsc{21cmFAST} \citep{2011MNRAS.411..955M} with UV background model \citep{2013MNRAS.432.3340S} implemented in the \textit{Meraxes} semi-analytic galaxy evolution model \citep{2016MNRAS.462..250M}. First, we construct total mass, stellar mass, and star formation rate grids from the particle distribution via a cloud-in-cell algorithm. A series of 
real-space spherical tophat filters are then applied to each grid, beginning from a radius of $R_{\mathrm{max}} = 20.34 h^{-1} \mathrm{cMpc}$, decreasing in radius by a factor of $\Delta R = 1.1$ down to the size of the grid cell. These smoothed grids represent the amount of mass, stars and star-formation within a sphere of radius R centred on the cell. The maximum filter radius $R_{\mathrm{max}}$ was chosen to be comparable to extrapolation of measurements of the mean free path of UV photons in an ionised medium at $z<5$ \citep{2010ApJ...721.1448S,Worseck14}, although the model does not include a physical prescription for the mean free path (see \citet{Davies22}). The interpretation of the maximum radius as a mean free path is inconsistent with recent measurements showing a rapid decrease at $z=6$ \citep{Becker21}. However it is unclear whether the cause of this decrease is driven by the structure of residual neutral hydrogen, self-shielding regions on small scales, or the epoch of reionisation being incomplete at $z=6$. Our maximum filter radius $R_\mathrm{max}$ is meant to represent a maximum scale which a single source can ionise, and is hence more analogous to the $z=5$ mean free path where we expect a much more ionised medium.

If at any filter radius, a cell in the filtered stellar grid $m_*$ and gas mass grid $M_{\mathrm{tot}}$ satisfies

\begin{equation}\label{eq:ioncondition}
    \zeta \frac{m_*}{M_{\mathrm{tot}}} > 1,
\end{equation}
where
\begin{equation}\label{eq:ionefficiency2}
    \zeta = \frac{N_\gamma}{f_b (1 - \frac{3}{4} Y_{\rm{He}})}
\end{equation}

\begin{equation}
    m_{*} = f_{10} \sum_i^{\mathrm{cell}} m_i \left( \frac{M_{\mathrm{halo},\it{i}}}{10^{10} M_\odot} \right)^{f_\alpha}
\end{equation}

Then the cell is considered ionised. The escape fraction normalisation $f_{10}$ and halo mass scaling $f_\alpha$ are used to weight each stellar particles' mass $m_i$ and based on the total mass of the halo in which each star particle resides $M_{\mathrm{halo},\it{i}}$ and represent the fraction of ionising photons which escape the host halo into the IGM. In equation \ref{eq:ionefficiency2}, $N_\gamma = 4000$, $f_b = \frac{\Omega_b}{\Omega_m} \approx 0.165$, and $Y_{\mathrm{He}}=0.24$ such that $\zeta \approx 29564$ prior to the escape fraction weighting. We set $f_{10} = 0.25$ and $f_\alpha = 0.5$ so that the global reionization history generated from the stellar population is consistent with electron optical depth measurements from \citet{Planck2018}, and completes shortly after $z=6$, in agreement with Lyman alpha forest measurements of the tail of reionization \citep{Qin21,Keating20,Nasir20,Bosman22}. The halo mass scaling of the escape fraction can be driven by different feedback processes in galaxies which clear out columns of gas \citep{Seiler19}, allowing ionising photons to escape. While simulations more commonly predict that on average, the escape fraction is smaller in larger halos \citep{Ferrara13,Paardekooper15,Xu16}, $f_{\alpha}$ is not well constrained by current observational data \citep{Park19,Mutch23}. In the case of \textsc{Astrid}, the stellar mass and UV luminosity functions sit $\sim 1 \sigma$ below the mean of the observations at redshifts $6<z<8$ (see \citet{Bird21}), particularly at low stellar masses. This stellar mass distribution implies a relatively high escape fraction which scales positively with halo mass, in order to match observations of global neutral fraction.

\begin{figure*}
    \centering
    \includegraphics[width=0.8\linewidth]{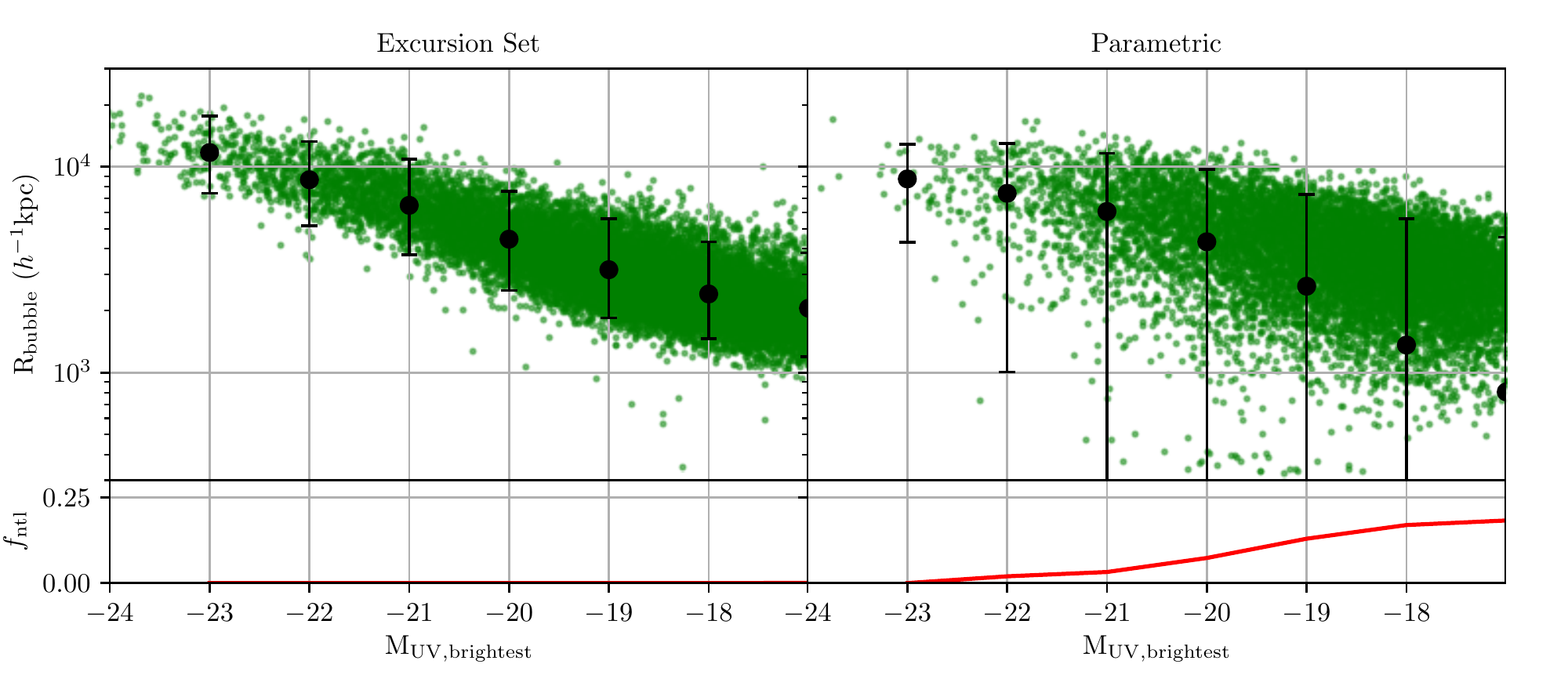}
    \caption{Ionised bubble radius versus the UV magnitude of the brightest galaxies within them in \textsc{Astrid-ES} (left) and \textsc{Astrid} (right). Each point represents one galaxy which is brightest in its HII region. Points with error bars represent the binned mean and $95^{\mathrm{th}}$ percentile range in bins of width 1. We present the snapshots in the left column of figure \ref{fig:reion50} where the universe is $\approx 22 \%$ ionised. Bubble radius is more closely correlated with UV magnitude in the excursion-set model, and there are many bright galaxies outside of ionised regions ($R_{\mathrm{bubble}} = 0$) as shown in the bottom panels, which show the fraction $f_\mathrm{ntl}$ of FOF groups (including those which aren't brightest in their region) within each bin that exists in a neutral region.}
    \label{fig:rvsm50}
\end{figure*}

Once the reionization state of the universe is calculated for a particular snapshot, we find the specific ionising intensity $J_{21}$ for each fully ionised cell, on the largest filter scale which satisfies the condition specified in equation \ref{eq:ioncondition}.

\begin{equation}
    J_{21} =  \frac{(1+z)^2}{4\pi} \lambda_{\mathrm{mfp}} h \alpha_{\mathrm{UV}} \bar{\epsilon} \times 10^{-21} \; \mathrm{erg} \, \mathrm{cm}^{-2} \mathrm{Hz}^{-1} \mathrm{s}^{-1} \mathrm{sr}^{-1},
\end{equation}
where $h$ is the Planck constant, $\lambda_{\mathrm{mfp}}$ is the mean-free path of ionising photons (here assumed to be the bubble radius and filter scale, $r$), and the mean emissivity in the sphere $\bar{\epsilon}$ is given by
\begin{equation}
    \bar{\epsilon} = \frac{N_\gamma}{\frac{4}{3} \pi r^3 m_p} \frac{\delta m_*}{\delta t}
\end{equation}
$\frac{\delta m_*}{\delta t}$ refers to the star formation rate within the cell, smoothed on the same scale as the stellar mass grid and weighted by the same escape fraction. Since the star formation rates of gas particles are very bursty, we smooth the star formation rate over a fraction of the Hubble time
\begin{equation}
    \frac{\delta m_*}{\delta t} = \frac{m_*}{t_* H^{-1}(t)}
\end{equation}
Where we set $t_* = 0.4$, and $H^{-1}(t)$ is the Hubble time. This smoothing is necessary in order to function with the cooling model in \textsc{MP-Gadget} which assumes ionisation equilibrium.
    
Local ionisation and heating rates can be calculated from $J_{21}$ assuming a spectral slope $\alpha_{\mathrm{UV}}$ above the HI ionisation threshold $\nu_{\mathrm{HI}}$ such that $J_\nu = J_{21} \left( \frac{\nu}{\nu_{\mathrm{HI}}} \right) ^{-\alpha_{\mathrm{UV}}}$, by integrating over the source spectra and species cross section $\sigma_i (\nu)$ above the ionisation threshold $\nu_i$ for each species $i = \mathrm{\{HI,HeI,HeII\}}$.

\begin{equation}\label{eq:ionrate2}
\Gamma _i = \int _{\nu _i} ^{\infty} \frac{4\pi J_\nu}{h_{\mathrm p} \nu} \sigma _i(\nu)d\nu \; \; \mathrm{s}^{-1},
\end{equation}
\begin{equation}\label{eq:heatrate2}
g_i = \int _{\nu _i} ^{\infty} \frac{4\pi J_\nu}{\nu}(\nu - \nu _i) \sigma _i(\nu)d\nu \; \; \mathrm{erg \, s}^{-1},
\end{equation}
where we use $\alpha_{\mathrm{UV}} = 2$ in the following simulations. The heating and ionisation rates are input to the existing cooling code (following \citet{Katz96} equations 23-32) which uses each $\Gamma _i$ to solve for an equilibrium ionisation state at each timestep and tracks changes in internal particle energy via the ionisation heating rates $g_i$ as well as recombinations, collisions, Bremsstrahlung, and inverse Compton cooling. The internal energy of each particle then feeds back onto processes such as gas cooling and star formation. The ionising background is calculated from the distribution of particles every $10$ Myr and assumed to be constant between updates.

When a region is first ionised, it is heated to a high temperature $> 10^4 K$ depending on the background ionising spectrum and the speed of the ionisation front \citep{DAloisio19}. Since our timesteps between excursion set calculations are too long to reliably calculate ionisation front speed, we set the post-reionisation temperature of each particle to $T_\mathrm{reion} = 15000 K$, corresponding to a scenario where ionisation fronts are slowed significantly by unresolved absorbers. $T_\mathrm{reion}$, the star formation rate timescale $t_*$, and the spectral slope of sources $\alpha_{\mathrm{UV}}$ are chosen in order to be simultaneously consistent with high-redshift IGM temperature measurements from \citet{Gaikwad20} and ionisation rate measurements \citep{Wyithe11,Calverley11,DAloisio18,Becker21}.


\section{Results}\label{sec:results}
We present reionization realizations from \textsc{Astrid} with the existing parametric reionization model on a $248^3$ grid, and from a re-run of reionisation from $z=20$ to $z=5.5$ labelled \textsc{Astrid-ES}, which includes the excursion-set model performed on a $500^3$ grid and described in section \ref{sec:es}. Both simulations have a dark matter particle mass of $6.75 \times 10^6 h^{-1} M_\odot$ and a gas particle mass of $1.267 \times 10^6 h^{-1} M_\odot$, with initial conditions containing $2 \times 5500^3$ particles within a cube of side length $250 h^{-1} \mathrm{Mpc}$. We show the galaxy stellar mass functions of \textsc{Astrid-ES} at redshifts $z=8$, $z=7$ and $z=6$ in Figure \ref{fig:gsmf}, which are in agreement with \textsc{Astrid} \citep{Bird21}.

Figure \ref{fig:globaleor} shows the global reionization history of \textsc{Astrid} and \textsc{Astrid-ES}, compared with the Planck 2018 electron optical depth measurements \citep{Planck2018} and observational constraints on $X_\mathrm{HI}$ from dark pixels \citep{McGreer15}, Lyman alpha emitter fraction \citep{Inoue18,Ouchi18,Morales21,Wold21}, lyman alpha equivalent width \citep{Hoag19,Mason19,Whitler20} and Quasar damping wings \citep{Greig17,Banados18,Davies18,Greig19,Wang20}. The parametric model reionises earlier, with a midpoint of $z \approx 7.5$, chosen to match the \citet{Planck2018} electron optical depth measurements. We also show that the volume-averaged ionisation rate $\Gamma_{12}$ and the temperature at mean density $T_0$ of the excursion set model are consistent with observations \citep{Wyithe11,Calverley11,DAloisio18,Becker21,Gaikwad20}. The late reionisation of the excursion-set model is consistent with inference on large-scale Lyman alpha optical depth fluctuations \citep{Qin21,Choudhury21,Bosman22} which suggest that the final overlap stages of reionisation $x_\mathrm{HI} \lesssim 5 \%$ should occur at late times $z<5.6$.

\subsection{Comparison of models}
We first present a comparison between the reionization topologies of the parametric reionization model in \textsc{Astrid} and the excursion-set algorithm in \textsc{Astrid-ES}. Figure \ref{fig:reion50} shows the reionization topology of the simulation between redshifts $z=8$ and $z=5.5$. When comparing these models, we select snapshots from each simulation that have a fraction of ionised cells (those satisfying the condition in equation \ref{eq:ioncondition}) that are approximately equal in order to present each simulation at a similar stage of cosmic reionization.

\begin{figure}
    \centering
    \includegraphics[width=\linewidth]{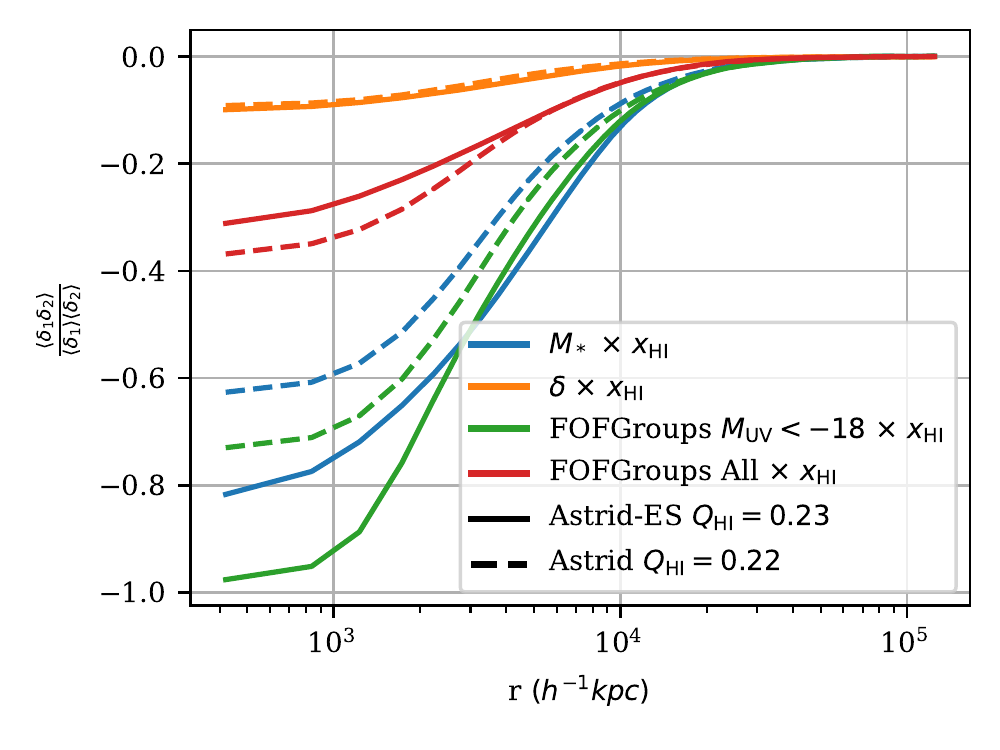}
    \caption{The cross-correlation functions between the $x_\mathrm{HI}$ field and other fields in each simulation at a neutral fraction of $Q_\mathrm{HI} \approx 0.22$, using a $0.5 \mathrm{Mpc}$ resolution. We show the cross-correlation of neutral fraction with density (orange), stellar mass (blue), All FOF groups (green) and FOF groups with a UV magnitude below -18 (red). Cross-correlations in \textsc{Astrid-ES} are shown with solid lines and those in \textsc{Astrid} are showin with dased lines. On all scales, the \textsc{Astrid-ES} ionization field is more strongly correlated with the stellar mass and brighter galaxies, whereas in \textsc{Astrid} there is has a slightly stronger correlation with the density field.}
    \label{fig:xcorr}
\end{figure}

As expected, the parametric model proceeds outward from the dense cosmic filaments, regardless of where stars form in the simulation. In the excursion-set algorithm, spherical ionised regions form around areas with high stellar mass. The highly spherical nature of the ionised regions in the excursion set arises from our choice of escape fraction which scales positively with halo mass, and is consistent with other reionization models driven by larger sources (e.g. \citet{McQuinn07,2016MNRAS.462..804G}). To show the connection between star formation and reionization quantitatively, we examine the cross-correlation between the density, stellar mass, and galaxy fields in Figure \ref{fig:xcorr}. The \textsc{Astrid-ES} ionisation field has a stronger negative correlation with the stellar mass, reaching $\sim -0.75$ at scales of $1 h^{-1}\mathrm{Mpc}$, compared to $\sim -0.6$ in \textsc{Astrid}. While the cross-correlation between all Friends-of-Friends groups and the ionization field is stronger in \textsc{Astrid}, this includes many low-mass structures which better trace the density field than the stellar mass field. Placing a UV Magnitude cut on the objects at $M_{UV} < -18$ returns an even stronger correlation in both simulations due to galaxy bias, however in \textsc{Astrid-ES} the increase is sharper, due to both the direct dependence of reionization history on stellar mass as well as the positive escape fraction weighting with halo mass, which increases the effect of the largest structures on reionization. The UV magnitude is calculated per Friends-of-Friends group from its star formation rate \citep{Stringer11}:

\begin{equation}
    \mathrm{M_{UV}} = -2.5\mathrm{log}_{10}(\psi) - 18.45,
\end{equation}

where $\psi$ is the star formation rate in solar masses per year.
We can also see the connection between galaxies and the ionised structure by comparing the size of the ionised regions to the UV magnitude of the brightest galaxies within them. We measure the relationship between galaxy UV magnitude and HII region size in Figure \ref{fig:rvsm50}, where the radius of each ionised region is calculated by taking the average of the distance in $5000$ random directions from a galaxy to an ionisation boundary $X_{\mathrm{HI}} > 0.5$, and eliminating galaxies that are not the brightest (lowest $\mathrm{M_{UV}}$) within their radius.

\textsc{Astrid-ES} has a much tighter correlation between ionised region size and the UV magnitude of their central galaxies. \textsc{Astrid} also has many bright galaxies outside of HII regions ($R=0$), including $\sim 10\%$ of galaxies at $\mathrm{M_{UV}} = -20$, up to $\sim 20\%$ of galaxies at $\mathrm{M_{UV}} = -17$. This is an effect of the parametric model not directly depending on galaxy properties, such that relatively bright galaxies outside of dense regions will not ionise their surroundings. However, it should be noted that our choice of escape fraction which scales positively with halo mass will also contribute to the tight correlation between $\mathrm{M_{UV}}$ and HII region size, since brighter galaxies will contribute a greater proportion of the ionising photons within each HII region.

One of the key observables during reionization is the 21cm power spectrum. This is calculated from the 21cm brightness temperature, which depends on the neutral fraction $x_{\mathrm{HI}}$ and overdensity $\delta$ grids. Here we use \citep{Furlanetto06}:

\begin{equation}\label{eq:dtb5}
    \delta T_{\mathrm{b,21}} = 27 \mathrm{K} \left( \frac{\Omega_b h^2}{0.023} \right)
    \left( \frac{0.15}{\Omega_m h^2}\frac{1+z}{10} \right) ^ {1/2}
    \left( 1-\frac{T_{\mathrm{CMB}}}{T_{\mathrm{s}}} \right) (1+\delta)x_{\mathrm{HI}}
\end{equation}

\begin{figure}
    \centering
    \includegraphics[width=\linewidth]{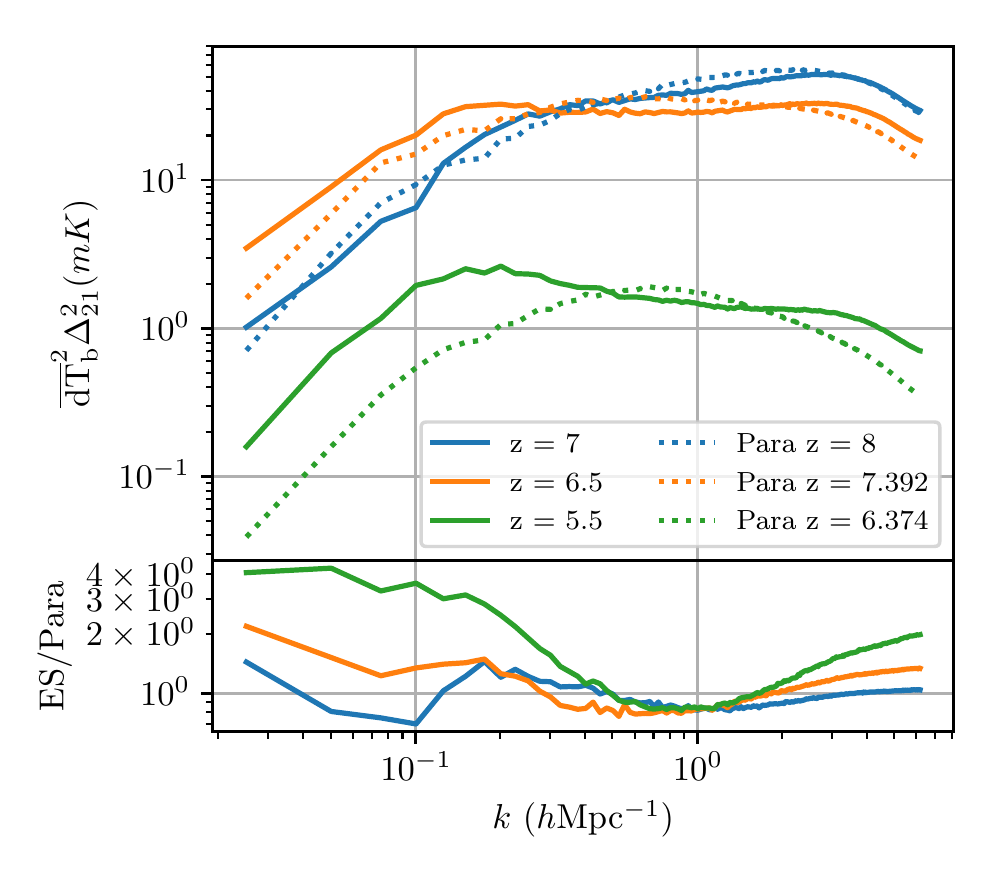}
    \caption{Top Panel: Dimensional power spectrum $\overline{\delta T}_{\mathrm{b,21}} (z)^2 \Delta_{21}^2$ in \textsc{Astrid-ES} (solid lines) and \textsc{Astrid} (dotted lines). Snapshots for each simulation excursion-set model are chosen at matching stages of reionisation, as in figure \ref{fig:reion50}, when the universe is $\approx 22 \%$, $50 \%$, and $95 \%$ ionised. Bottom Panel: Ratio of the power spectra in the excursion set and parametric models at the matched snapshots.}
    \label{fig:ps50}
\end{figure}

We assumed the spin temperature of the gas is saturated $T_{\mathrm{s}} >> T_{\mathrm{CMB}}$, which is valid in the scenario where X-ray heating from early quasars occurs prior to the bulk of reionization \citep{Furlanetto06}. This formula also ignores redshift space distortions due to the peculiar velocities of the gas.

The spherically averaged 21cm power spectrum is then calculated from the Fourier transform of the brightness temperature grid on a scale $k$ (e.g \citep{Lidz08}):

\begin{equation}
    \Delta_{21}^2 (k,z) = \frac{k^3}{2\pi V} \left< |\delta_{21}(k,z)|^2_k \right>,
\end{equation}
where
\begin{equation}
    \delta_{21}(k,z) = \frac{\delta T_{\mathrm{b,21}}(k,z)}{\overline{\delta T}_{\mathrm{b,21}}(z)} - 1
\end{equation}
and $\overline{\delta T}_{\mathrm{b,21}}(z)$ is the average brightness temperature in the simulation at redshift $z$, and $V$ is the simulation volume. Due to the difficulty in measuring the 21cm signal from the EoR, only upper limits have been placed on the 21cm power spectrum so far \citep{Trott20,Mertens20,Patil17}. However, the 21cm power spectrum contains a lot of information on the scale of ionised and neutral regions during reionization, and will be one of the most important observables used to constrain our models in the near future \citep{Greig20MWA,Greig21LOFAR}.

We show the resulting power spectra in Figure \ref{fig:ps50}. There are two main differences in the power-spectra between the excursion-set and parametric reionisation models. Firstly, the excursion set contains fewer and larger ionised regions at a fixed $X_{\mathrm{HI}}$, which increases large scale power, especially toward the end of reionisation where the last remaining neutral regions are much larger. This is caused by a combination of the escape fraction scaling and the clustering of stellar particles compared to the density field. Secondly, the more spherical ionised regions change the shape of the power-spectrum, showing a slight increase in power on scales $0.1 \lesssim k \lesssim 0.5 h\mathrm{Mpc}^{-1}$, corresponding to the size of the ionised regions $5 \lesssim R \lesssim 30 h^{-1}\mathrm{Mpc}$. The more asymmetric shape of ionised regions in the parametric model spreads this power over a larger range of $k$.


\begin{figure*}
    \centering
    \includegraphics[width=0.9\linewidth]{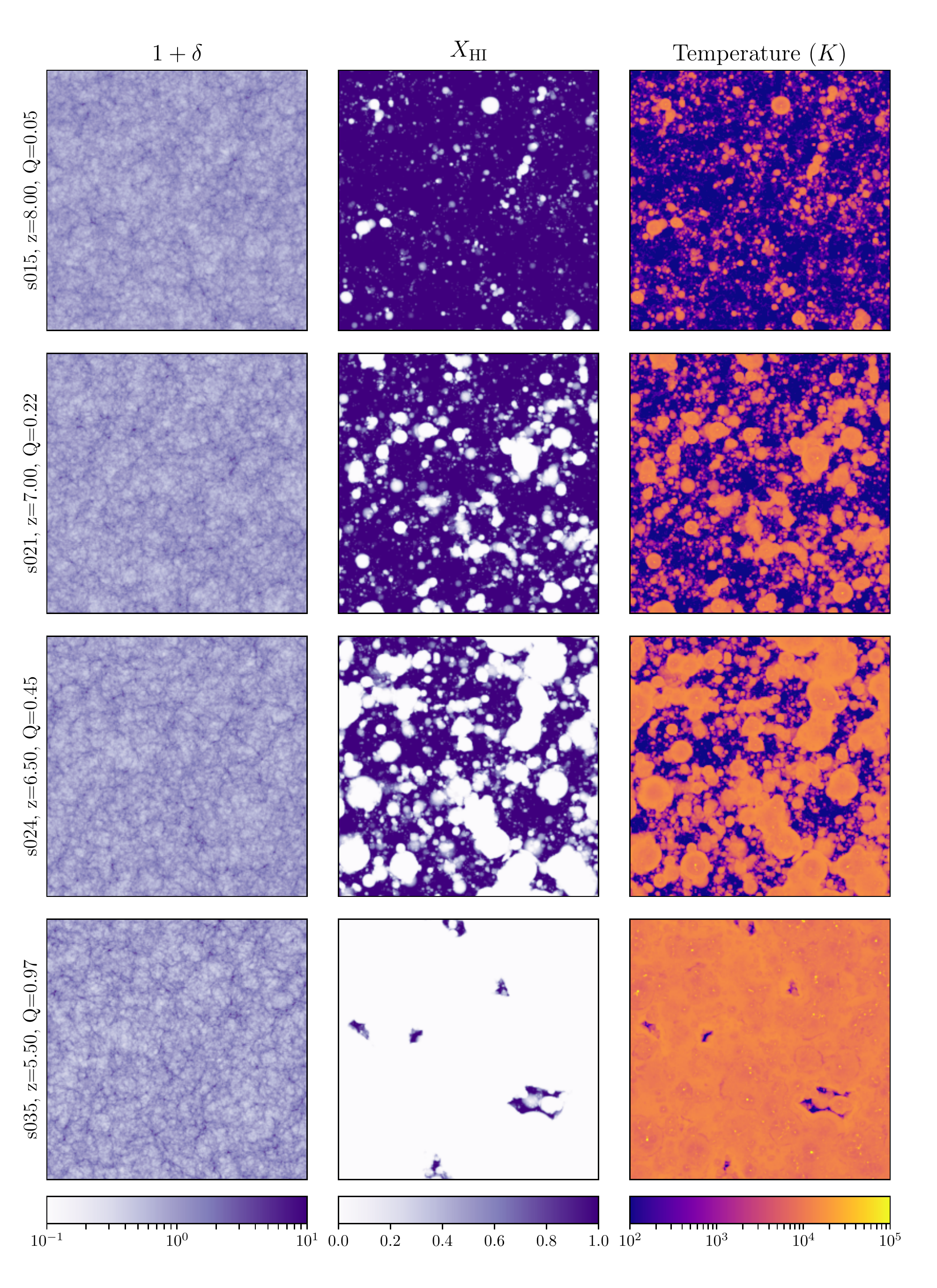}
    \caption{$250 \times 250 \times 5 h^{-1}\mathrm{Mpc}$ slices of\textsc{Astrid-ES} between redshifts $z=8$ and $z=5.5$. Left: Matter overdensity, Centre: Neutral hydrogen fraction. Right: Gas temperature. Spherical ionised regions form and expand from the dense filaments of the cosmic web, heating the gas to $T = 1.5 \times 10^4 \mathrm{K}$.}
    \label{fig:reion100}
\end{figure*}

\begin{figure*}
    \centering
    \includegraphics[width=\linewidth]{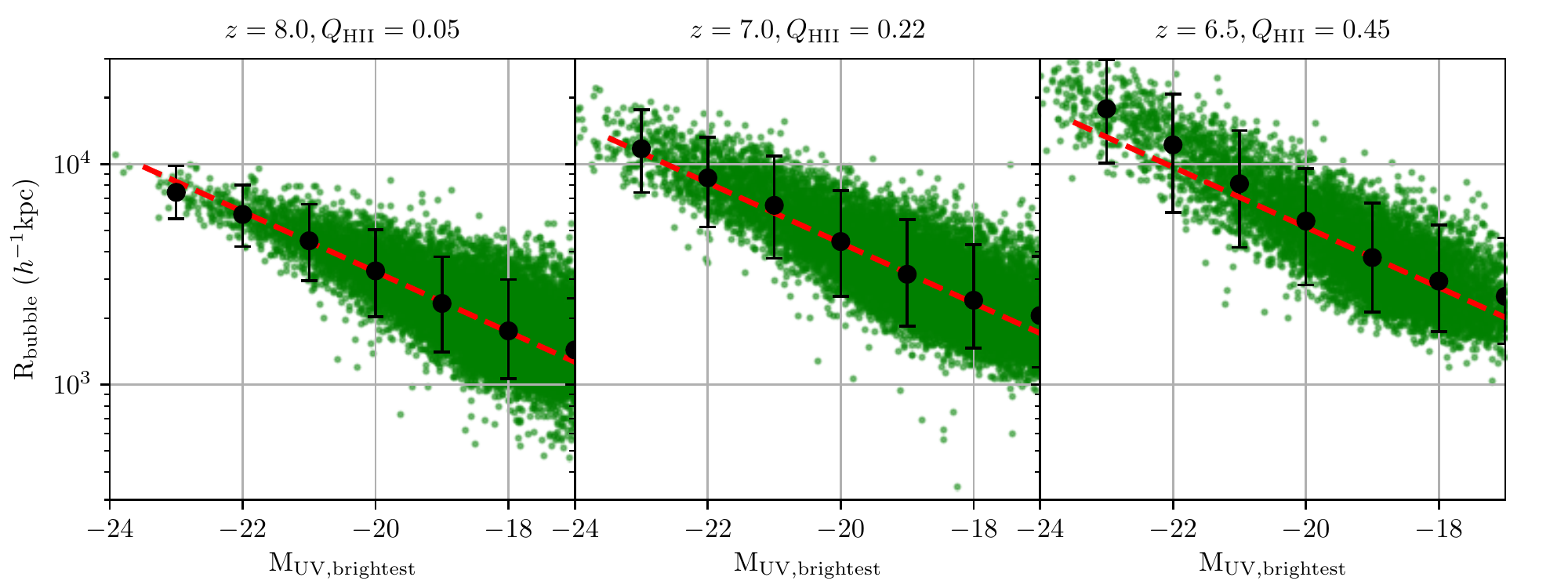}
    \caption{HII region radius versus the UV magnitude of the brightest galaxy within, at redshifts $z=8.5$, $z=7$, and $z=6.5$ in \textsc{Astrid-ES}. Points with error bars represent the mean and $95^{\mathrm{th}}$ percentile range in radius, within bins of width 1. The red dashed line shows a power law fit to this relation $\mathrm{log} (R) = -0.314 M_\mathrm{UV} - 2.550 \mathrm{log}(1+z) + 7.408$ using the snapshots at $z=8.0$, $z=7.5$, $z=7.0$, and $z=6.5$.}
    \label{fig:rvsm100}
\end{figure*}

\subsection{Excursion-Set Results}
In this section we present more detailed results for \textsc{Astrid-ES}. Figure \ref{fig:reion100} shows the reionization history of a $250 \times 250 \times 5 h^{-1} \mathrm{Mpc}$ slice of the simulation, showing the neutral fraction, matter overdensity, and gas temperature at a resolution of $0.5 h^{-1} \mathrm{Mpc}$. The first ionised regions grow in the overdense regions of the simulation where most of the stars form. Reionization proceeds outward from these regions as new bubbles form, expand and overlap to cover the entire volume shortly after redshift $z=6$. As the ionisation front passes, gas is heated to $15000 \mathrm{K}$ (see section \ref{sec:es}). While we stop the simulation at the last stages of reionisation, we can see some of the earliest ionised regions begin to cool toward a temperature closely correlated with their density \citep{2016MNRAS.460.1885U}.

We show the correlation between HII region size and UV magnitude at redshifts $z=8$, $z=7$, and $z=6.5$ for \textsc{Astrid-ES} in Figure \ref{fig:rvsm100}. We find that a power-law fit of $\mathrm{log} (R) = -0.314 M_\mathrm{UV} - 2.550 \mathrm{log}(1+z) + 7.408$ well describes the data pre-overlap, with a standard deviation of $\sigma_R < 0.15 \mathrm{dex}$ across all mass bins and redshifts. Since brightest galaxies within a HII region are selected by excluding galaxies which have a neighbor within $\mathrm{R}$ with a lower $M_\mathrm{UV}$, the fit becomes less accurate during overlap, where HII regions are less spherical, and dimmer galaxies within a larger HII structure are included in the sample.

\begin{figure}
    \centering
    \includegraphics[width=\linewidth]{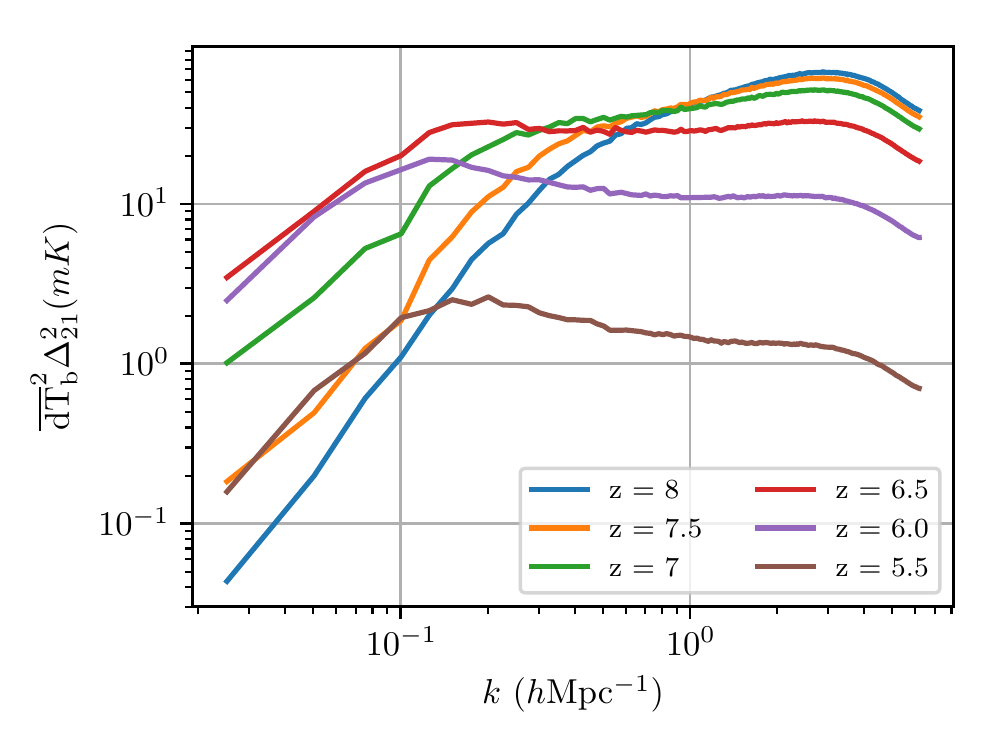}
    \caption{The dimensional power spectrum $\overline{\delta T}_{\mathrm{b,21}} (z)^2 \Delta_{21}^2$ in \textsc{Astrid-ES} between redshifts $z=8$ and $z=6$. The power spectrum shape initially follows the density field, power on large scales then increases as ionised regions form and grow. Once reionization nears completion, power on all scales decreases as the 21cm brightness temperature approaches zero.}
    \label{fig:PS_100}
\end{figure}

We present the dimensional 21cm power spectrum $\overline{\delta T}_{\mathrm{b,21}} (z)^2 \Delta_{21}^2$ throughout reionization in our simulation in Figure \ref{fig:PS_100}. While the universe is entirely neutral, the shape of the 21cm power spectrum resembles the matter power spectrum. Once the first ionised regions start to form, power increases on large scales $k \lesssim 1 h\mathrm{Mpc}^{-1}$, corresponding to the size of the bubbles. Once the universe is mostly reionised, the power on all scales decreases as the 21cm signal from atomic hydrogen disappears. The qualitative shape of the power spectrum is similar to many reionization models \citep{Lidz08}, but the increase in power on large scales, which results from the distribution in HII region sizes, will take a different form depending on factors such as escape fraction dependencies and feedback processes. This means that predictions for the 21cm power spectrum can be useful in ruling out certain reionization models \citep{2016MNRAS.462..804G,Greig20MWA,Greig21LOFAR}.

\subsection{Reionization Feedback}\label{sec:feedback}
The photo-ionisation of intergalactic hydrogen is accompanied by rapid heating, as the excess energy of ionising photons is converted to kinetic energy in the IGM \citep{1994MNRAS.266..343M,1997MNRAS.292...27H}. This heating will affect star-formation in lower mass galaxies, as gas from the IGM is less likely to cool toward the centre of a halo \citep{Couchman86,Sobacchi13a}. Other works have found that while reionization feedback does not greatly affect the progress of reionization compared to other feedback mechanisms, it significantly decreases star formation in small halos over time \citep{2013MNRAS.432.3340S,2016MNRAS.462..250M,Hutter21}.

\begin{figure*}
    \centering
    \includegraphics[width=\linewidth]{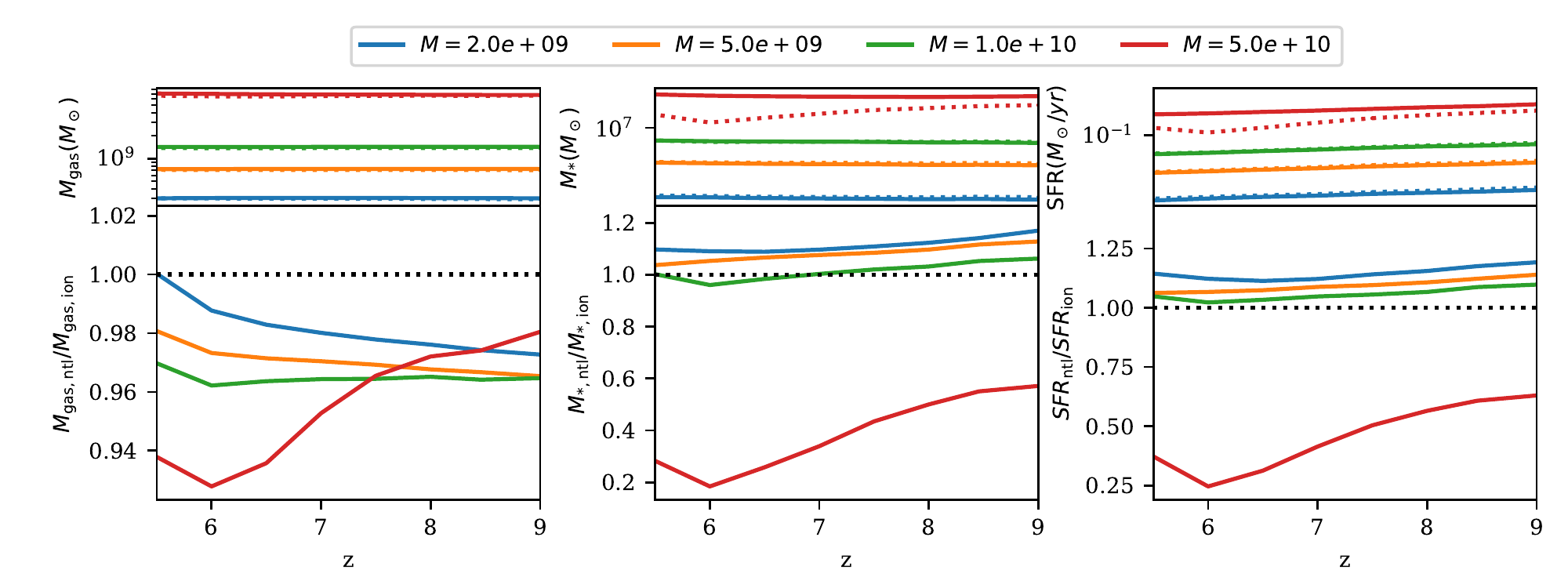}
    \caption{Top: Mean gas mass (left), stellar mass (middle) and star formation rate (right) in Star formation rate distributions in \textsc{Astrid-ES} between redshifts $5 < z < 9$. The mean SFR for halos within large-scale neutral regions are shown as dotted lines, and for those within ionised within ionised regions as solid lines. The distributions are calculated on halo mass bins within $30\%$ of $2\times10^9$, $5\times10^9$, $1\times10^{10}$, and $5\times10^{10}$ solar masses. Bottom: Ratio of each halo property in neutral regions to ionised regions, at the same masses and redshifts.}
    \label{fig:fofz}
\end{figure*}

In \textsc{Astrid}, \citet{Bird21} found up to a $50\%$ reduction in star formation in within $2 \times 10^9 M_\odot$ halos which reside in large-scale ionised regions compared to halos in neutral regions, toward the end of reionization. On the other hand, large halos had similar SFR regardless of whether or not they are in an ionised region. Since the UV background in \textsc{Astrid-ES} is included on-the-fly and results directly from the sources in the simulation, we repeat and expand on this analysis to determine if these correlations persist in a more self-consistent reionisation model. The UV background given by the excursion set (equations \ref{eq:ionrate2} - \ref{eq:heatrate2}) is applied to the gas particles in an ionised cell, as well as the sharp temperature increase of $T_{\mathrm{reion}} = 15000K$ at the time of reionisation. This heating can reduce the amount of cold gas available for star formation, particularly in less massive halos.

We examine the same results in \textsc{Astrid-ES} in Figure \ref{fig:fofz}, additionally including gas mass and stellar mass in halos. The correlation between star formation rates and ionisation state is similar to \textsc{Astrid} in the lowest masses, but is weaker toward higher masses, and reversed in the highest mass bin with $\sim 5 \times 10^{10} M_\odot$ halos having mean star formation rates up to $75\%$ lower in neutral regions. This is due to the direct connection between the star formation in these groups and reionization structure in the excursion-set model, and is likely exaggerated by our increasing escape fraction toward larger halos. In order for a halo of this size to exist in a neutral region, it \textit{must} have a relatively low star formation, otherwise it would have ionised its surroundings.
The median stellar mass follows a similar pattern, being slightly higher in lower mass halos, and much lower in high-mass halos within neutral regions. The median gas mass shows the opposite trend at low masses, being lower in neutral regions at all masses.

The differences between these properties in ionised and neutral regions of the universe can be driven by a number of effects. The heated gas left behind after reionization may not cool as quickly onto halos or form stars. This means that existing halos will form stars more slowly, and newly formed halos will contain fewer stars than those forming in neutral regions. This drives an anti-correlation between star formation and ionisation for smaller halos with shallower potential wells, as well as at later times when halos have spent more time on average surrounded by a heated IGM. Highly star forming galaxies are also more likely to ionise their surroundings, and therefore exist within ionised regions, causing a strong correlation between star formation and ionisation at higher masses while reionisation is ongoing.

\begin{figure}
    \centering
    \includegraphics[width=\linewidth]{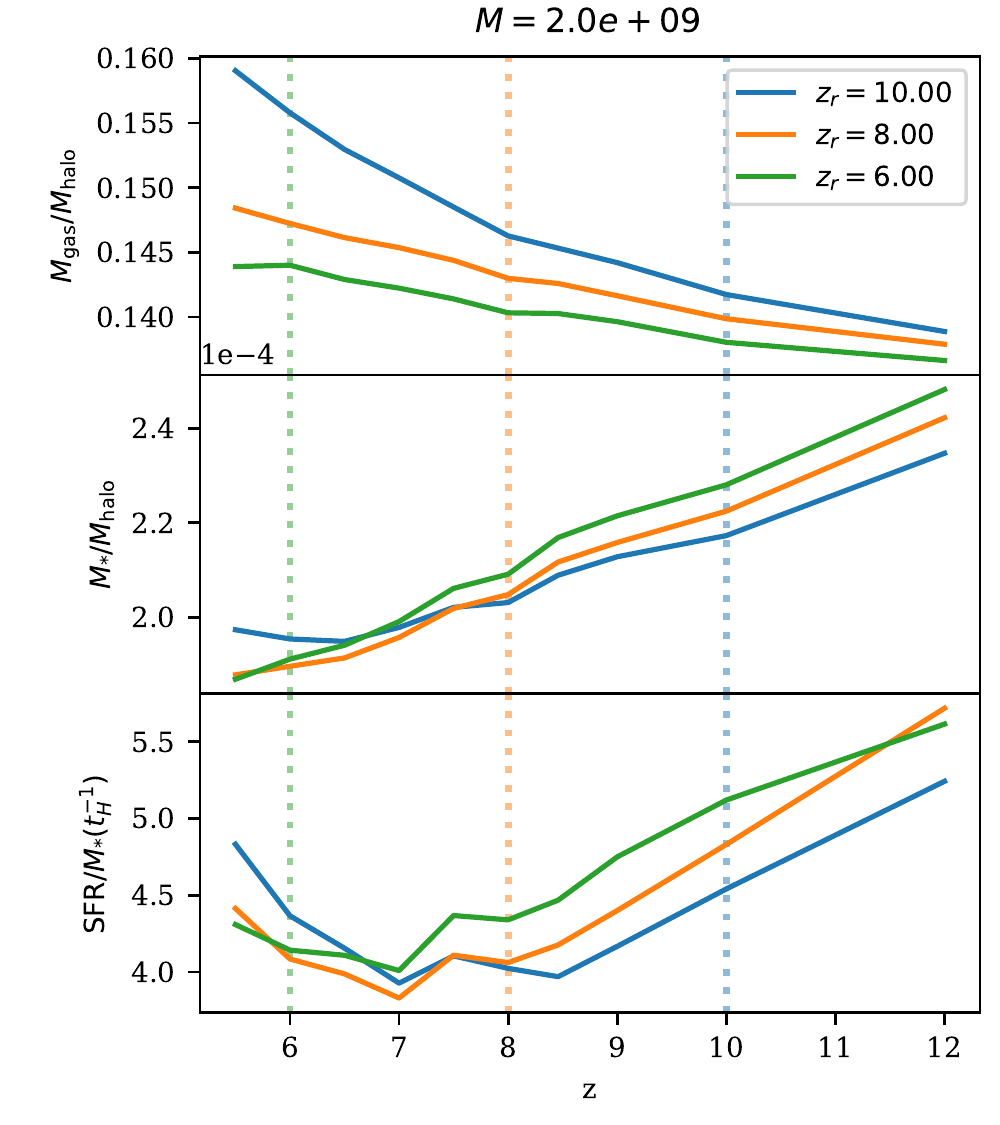}
    \caption{Mean halo gas mass (Top), stellar mass fraction (Middle) and specific star formation rate (Bottom) in halos of mass $M_\mathrm{halo} \approx 2 \times 10^9 h^{-1}M_\odot$ which ionise around $z=10$ (blue line), $z=8$ (orange line), and $z=6$ (green line). The reionisation timing of each group is marked with a vertical dotted line of the corresponding colour. While there are correlations between halo properties and reionisation timing at fixed halo mass, there is no sudden change or acceleration in each property at the redshift of reionisation, meaning that reionisation feedback likely plays a subdominant role in the evolution of these properties.}
    \label{fig:fofdz}
\end{figure}

In order to examine these effects more closely, we focus on the lowest mass bin $M_\mathrm{halo} \approx 2.0 \times 10^9$ solar masses, which we expect to experience the largest effect from reionisation feedback. We plot the median gas mass fraction, stellar mass fraction, and specific star formation rate for groups of halos which ionised at redshifts $z \approx 10$, $z \approx 8$, and $z \approx 6$ in figure \ref{fig:fofdz}. The $z_\mathrm{reion} \approx 10$ halos contain up to $\sim 5 \%$ more gas, $\sim 5 \%$ less stellar mass, and $\sim 10 \%$ less star formation. Suggesting that low mass halos which ionise earlier form stars from their gas reservoirs more slowly. However, these correlations persist both before and after the IGM surrounding the halos ionise, and do not show a sudden change or acceleration post-ionisation. This suggests that the less efficient star formation does not result from reionisation feedback, and that the heating of the IGM on large scales by reionisation does not greatly affect the cooling of gas onto halos within the timeframe of our simulation. Instead, it is likely that the differences come from environmental effects, where the regions that form the first ionising sources at higher masses form stars more slowly in lower mass halos. Since we expect reionization feedback to have a cumulative effect on galaxies, running our simulation to lower redshift may reveal a larger effect as halos are exposed to the UV background for longer. It is also worth mentioning that the reionisation temperature $T_\mathrm{reion} = 15000K$ chosen to match observations of IGM temperature at mean density is low compared to many models \citep{DAloisio18}, a higher reionisation temperature may also show a stronger effect on small halos. Comparing similar halos over a longer period of time, or matching halos across simulations with different reionisation histories (such as \textsc{Astrid} and \textsc{Astrid-ES}) would provide a more thorough study of feedback and environmental effects, however we leave this analysis to future work.

\section{Conclusion}
In this paper we have introduced the addition of an excursion-set reionization model based on \textsc{21cmFAST} into the hydrodynamic code \textsc{MP-Gadget}. This model is efficient enough to run on large simulations such as \textsc{Astrid} without significantly impacting the computational cost. We have performed a re-run of the epoch of reionization $5.5 < z < 20$ in \textsc{Astrid} labeled \textsc{Astrid-ES}. The global history, ionisation rates and mean temperatures in \textsc{Astrid-ES} are consistent with high-redshift quasar spectra and CMB measurements. We find that the excursion-set model correlates more strongly with the ionising sources in the simulation when compared against against the parametric model applied in the \textsc{Astrid} production run, in which the redshift of reionisation is calculated based on the initial mass density grid. With our choice of escape fraction, reionization in the excursion-set model contains fewer, larger ionised regions, which is reflected in the 21cm power spectrum. The model also includes the effect of reionization feedback, where an ionising background dependent on the galaxy star formation rate is applied to each particle. While there are significant correlations between halo properties and reionisation timing, they are likely driven by environmental effects, rather than by reionisation feedback, since they exist before and after the reionisation of the halo.

Future large simulations run with \textsc{MP-Gadget} will have the option to enable the excursion-set model, allowing for more accurate reionization topology and for further studies of the connection between galaxy properties and reionization. With a larger simulation, analyses of 21cm cross correlations \citep{Kubota18,Vrbanec20,Cox22} or predictions for 21cm images from the SKA \citep{Ghara20,Davies21} can be performed using a more accurate reionization model, producing more accurate predictions for the sensitivity of future observations to the parameters of reionization.

The source-driven reionization model will also allow the addition of further physics to the model, so that a wider range of observations can be predicted and calibrated against. The size of simulations such as \textsc{Astrid}, makes them ideal for studies of large scale X-ray heating during cosmic dawn \citep{Greig17heat}. A future work will include the X-ray heating of the IGM from galaxies, allowing us to make predictions for the spin temperature, $T_s$, include the effects on the EoR of early photoheating, and extend the range of predictions to include 21cm measurements toward higher redshifts, such as the data obtained by global 21cm experiments \citep{Bowman18,SinghSARAS}.

The large box size and black hole formation model within \textsc{Astrid} also makes it ideal for the study of the contribution of quasars to reionisation. The excursion set model can be extended to include quasars (e.g: \citet{Qin17QSO}). While most works find that quasars have a subdominant contribution to hydrogen reionisation \citep{Trebitsch23Astraeus}, their inclusion into the model would allow us to study the effect of rare, bright sources on the topology of reionisation \citep{Chardin17,Eide20}, and early quasar proximity zones \citep{Chen21CROC}. This would also provide a smooth coupling to the existing helium reionisation model within \textsc{MP-GADGET}, which will be useful to study the post-reionisation UV background.


\section*{Acknowledgements}
JD acknowledges support from the Ministry of Universities and Research (MUR) through the PRO3 project “Data Science methods for Multi-Messenger Astrophysics \& Multi-Survey Cosmology”.
This research was supported by the Australian Research Council Centre of Excellence for All Sky Astrophysics in 3 Dimensions (ASTRO 3D), through project number CE170100013.
This work utilised the
OzSTAR national facility at Swinburne University of Technology. 
OzSTAR is funded by Swinburne University of Technology and the
National Collaborative Research Infrastructure Strategy (NCRIS).
SB acknowledges funding support from NASA-80NSSC21K1840. The authors acknowledge the Frontera computing project at the Texas Advanced
Computing Center (TACC) for providing HPC and storage resources
that have contributed to the research results reported within this paper. Frontera is made possible by National Science Foundation award
OAC-1818253. URL: http://www.tacc.utexas.edu.

\section*{Data Availability}
The data underlying this article will be shared on reasonable request to the corresponding author. E-mail: james.davies@sns.it



\bibliographystyle{mnras}
\bibliography{paper.bib} 





\bsp	
\label{lastpage}
\end{document}